\documentstyle[prb,aps,amsfonts,amssymb,floats,epsfig]{revtex}
\begin{document}
\draft %
\twocolumn[\hsize\textwidth\columnwidth\hsize\csname
@twocolumnfalse\endcsname %
\title{Optical properties and electronic structure of
$\alpha^{\prime}$-Na$_{1-x}$Ca$_x$V$_2$O$_5$}
\author{C.~Presura$^1$, D. van der ~Marel$^1$, M.~Dischner$^2$, C.~Geibel$^2$, R.K.~Kremer$^3$}
\address{ $^1$Solid State Physics Laboratory, University of Groningen,
 Nijenborgh 4, 9747 AG Groningen, The Netherlands\\
 $^2$Max-Planck-Institut f\"{u}r Chemische Physik fester
  Stoffe, D-01187 Dresden, Germany\\
  $^3$Max-Planck-Institut f\"{u}r Festk\"{o}rperforschung,
  Heisenbergstrasse 1, 70569 Stuttgart, Germany}
\date{\today}
\maketitle
\begin{abstract}
The dielectric function of
$\alpha^{\prime}$-Na$_{1-x}$Ca$_x$V$_2$O$_5$ ($0 \le x \le 20 \%$)
was measured for the {\em a} and {\em b} axes in the photon energy
range 0.8-4.5 eV at room temperature. By varying the
Ca-concentration we control the relative abundancy of V$^{4+}$ and
V$^{5+}$. We observe that the intensity of the main optical
absorption peak at 1 eV is proportional to the number of
V$^{5+}$-ions. This rules out the interpretation as a V$^{4+}$
$d-d$ excitation, and it establishes that this is the on-rung
bonding-antibonding transition.
\end{abstract}
\vskip2pc]
\narrowtext
\section{Introduction}
$\alpha^{\prime}$-Na$_{1-x}$Ca$_x$V$_2$O$_5$ belongs to the larger
group of $\alpha^{\prime}$ vanadium pentoxides, with the chemical
formula AV$_2$O$_5$ (A= Li, Na, Ca, Mg, etc.) \cite {Galy}. Their
structure is remarkably similar to that of the parent compound
V$_2$O$_5$, which consists of layers of square pyramids of O
surrounding a V$^{+5}$ ion. The layers are kept together via weak
forces, which account for the easy cleavage of this oxide along
(001). The basic building blocks forming the V$_2$O$_5$ layers are
parallel ladders of VO$_5$ pyramids. In AV$_2$O$_5$ the A atoms
enter the space between the layers and act as electron donors for
the V$_2$O$_5$ layers. In the case of
$\alpha^{\prime}$-NaV$_2$O$_5$, every doped electron is shared
between two V atoms. As a result the average valence of the V-ions
corresponds to V$^{+4.5}$. X-ray diffraction
indicates that at room temperature all V-ions are in the same
mixed valence state \cite{meetsma,schnering,smolinski}.

Partial substitution of the Na$^+$ with Ca$^{2+}$ leaves the
$\alpha^{\prime}$ crystal structure intact, but alters the
relative abundance of V$^{4+}$ and V$^{5+}$,
$N^{4+}:N^{5+}=(1+x):(1-x)$. In this paper we report spectroscopic
ellipsometry measurements on
$\alpha^{\prime}$-Na$_{1-x}$Ca$_x$V$_2$O$_5$ (x=0, 0.06, 0.15 and
0.20), in the energy range 0.8-4.5 eV. We employ the dependence of
the optical spectra on the $N^{4+}:N^{5+}$ ratio to identify the
main components in the optical spectra, which in turn we use to
reveal the electronic structure of this material.
\section{Details of sample preparation and experimental setup}
The crystals (CR8, 45008, and 45010) had dimensions of
approximately 2, 1 and 0.3 mm along the {\em a}, {\em  b}, and
{\em c} axes respectively. The samples 45008 and 45010 were
prepared from NaVO$_3$ flux \cite{isobe}. In a first step a
mixture of Na$_2$CO$_3$ and V$_2$O$_5$ is heated up to 550¬ C in
air to form NaVO$_3$. In a second step the NaVO$_3$ is mixed with
VO$_2$ in the ratio of 8:1 and then heated up to 800¬ C in an
evacuated quartz tube and cooled down at a rate of 1\,K per hour.
The excess NaVO$_3$ was dissolved in water. Then the doped samples
were produced by substituting in the first step Na$_2$CO$_3$ by
CaCO$_3$. The chemical composition of the samples has been
determined using Energy Dispersive X-ray Fluoresence 
microprobe  measurements. The results showed that
the real Ca content of some samples was smaller that the nominal
one (with a factor of 0.75), and that position dependent
variations of the Na stoichiometry are below 2\%. A standard
spectroscopic ellipsometer was used to collect ellipsometric
data in the range of 6000 to 35000 cm${-1}$ from the $ab$ 
planes of the crystals using two different crystal orientations, 
and to measure normal incidence reflectivity spectra of the $ac$ 
plane with the electric field vector along the $c$-direction.
\section{Data collection and analyzes}
\begin{figure}
\centerline{\epsfig{figure=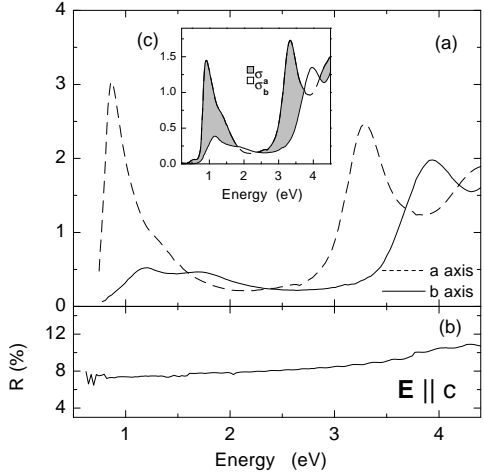,width=7cm, clip=}}
\caption{
(a) Pseudo-optical conductivity $\sigma_1^{ps}(\omega)$ of
$\alpha^{\prime}$-NaV$_2$O$_5$ at $T = 300$K \cite{note1}. The spectra were
taken on the (001) surface with $a$ and $b$ axes successively
lying in the plane of incidence. (b)  $E \| c$
reflectivity at $T = 300$K. (c) Optical conductivity
$\sigma_{a}(\omega)$ and $\sigma_{b}(\omega)$ corrected for
contributions of the c-axis dielectric function to
$\sigma_1^{ps}(\omega)$.}
\label{navops}
\end{figure}

We performed ellipsometric measurements on the (001) surfaces of
the crystals both with the plane of incidence of the light along
the {\em  a} and the {\em  b} axis. An angle of incidence
$\theta$, of $66^{0}$, was used in all experiments. Ellipsometry
provides directly the amplitude and phase of the ratio of the
reflectivity coefficients of $p$- and $s$-polarized light
\cite{azzam} $r_p(\omega)/r_s(\omega)$. For an anisotropic crystal
with the three optical axes ariented along the surface normal
($p\perp$), perpendicular to the plane of incidence (s), and along
the intersection of the plane of incidence and the surface
($p\parallel$), this ratio is related to the dielectric tensor
elements along these three directions ($\epsilon_{p\perp}$,
$\epsilon_{s}$, and $\epsilon_{p\parallel}$) according to the
expression:
\begin{equation}
 \begin{array}{l}
 \frac{r_p}{r_s} =
 \frac{\left[\sqrt{\epsilon_{p\parallel}\epsilon_{p\perp}}\cos\theta
  -\sqrt{\epsilon_{p\perp}-\sin^2\theta}\right]
 \left[\cos\theta+\sqrt{\epsilon_{s}-\sin^2\theta}\right]}
 {\left[\sqrt{\epsilon_{p\parallel}\epsilon_{p\perp}}\cos\theta
  +\sqrt{\epsilon_{p\perp}-\sin^2\theta}\right]
 \left[\cos\theta-\sqrt{\epsilon_{s}-\sin^2\theta}\right]}
 \label{ellips}
 \end{array}
\end{equation}
To extract the dielectric constant from
the ellipsometric parameters we proceed in two steps: First
the pseudo-dielectric functions along the optical axes
are extracted from the ellipsometric data using the
inversion formula
\begin{equation}
 \epsilon_{p\parallel}^{ps} =
  \sin^2\theta\left(1+\tan^2\theta\left(\frac{1-r_p/r_s}{1+r_p/r_s}\right)^2\right)
 \label{pseudo}
\end{equation}
For isotropic crystals this expression provides the dielectric
function directly. The pseudo-dielectric function is close to the
dielectric tensor elements along the intersection of the plane of
incidence and the crystal surface\cite{aspnes}. A biaxial crystal
like $\alpha^{\prime}$-NaV$_2$O$_5$ has three complex dielectric
functions, $\epsilon_a$, $\epsilon_b$ and $\epsilon_c$ along each
optical axis, and an ellipsometric measurement involves all three
tensor components of the dielectric matrix. In addition to the
pseudo-dielectric functions displayed  Fig.\ref{navops}a,
$\epsilon_c(\omega)$ is required. No $ac$ or $bc$ crystal-planes
were available large enough to do ellipsometry with our setup. We
therefore measured the $c$-axis reflectivity (Fig. \ref{navops}b)
of the $bc$-plane of the pristine material (sample CR3). The
spectrum contains no (or very weak) absorption peaks in the
measured frequency range, as it was reported earlier
\cite{konstantinovic,smirnov}, providing a very reliable
determination of the dielectric function $\epsilon_c$ using
Kramers-Kronig analysis. Due to the absence of strong resonances,
$\epsilon_c$ has a minor influence on the recorded ellipsometric
spectra. In Fig. \ref{navops}c the optical conductivity is
displayed taking into account all corrections due to the
anisotropy. We see that the conversion from
$\epsilon^{ps}(\omega)$ to $\epsilon(\omega)$ has indeed a rather
small effect on the spectra. In essence it leads to a factor 0.5
re-scaling of $\sigma(\omega)$.
\begin{figure}
\centerline{\epsfig{figure=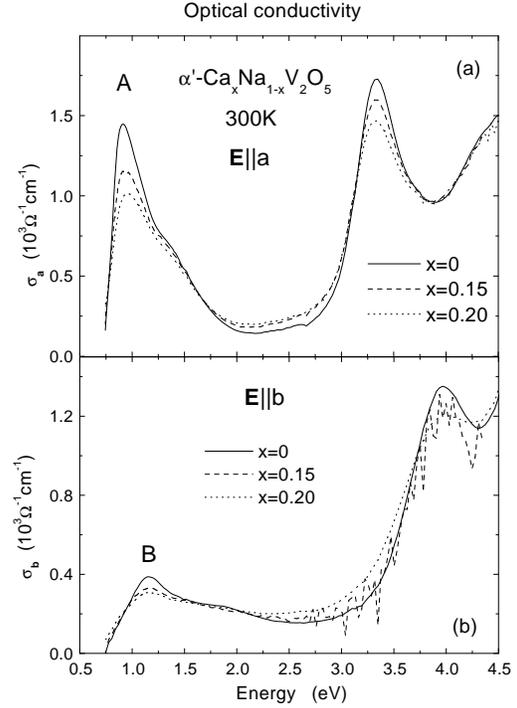,width=7cm, clip=}} \caption{
Optical conductivity at $T = 300$K of $\mbox{Re }\sigma(\omega)$
of Ca$_x$Na$_{1-x}$V$_2$O$_5$ for x=0 (solid), x=0.15 (dash) and
x=0.20 (dot): $E\parallel a$ (panel a) and $E\parallel b$ (panel
b)} \label{navoca}
\end{figure}
\begin{figure}
\centerline{\epsfig{figure=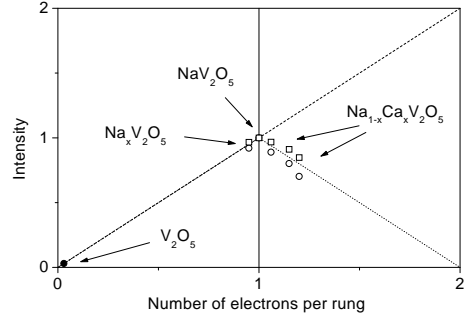,width=7cm, clip=}}
\caption{ Intensity of peak A calculated by integrating optical
conductivity up to 1.9eV (squares), or by using the height of the
peak (circles) in Ca$_x$Na$_{1-x}$V$_2$O$_5$ for x=0.06, x=0.15
and x=20 and in Na$_{x}$V$_2$O$_5$ for x=0.95. The dashed line
indicates the theoretical intensity of in-site V$dd$ transitions
versus doping. The dotted line indicates the theoretical intensity
according to the model Hamiltonian Eq.\ref{hamil}. }
\label{intens}
\end{figure}
\\
The data are in general agreement
with previous results \cite{damascelli,golubchik} using
Kramers-Kroning analysis of reflectivity data. Along
the a-direction we observe a peak at 0.9 eV with a shoulder
at 1.4 eV, a peak at 3.3 eV and the slope of a peak above 4.2 eV,
outside our spectral window. A similar blue-shifted sequence is
observed along the b-direction. The 1eV peak drops rather sharply
and extrapolates to zero at 0.7 eV. However, weak absorption has
been observed within the entire far and mid-infrared
range\cite{smirnov,damascelli}. The strong optical absorbtion
within the entire visible spectrum causes the characteristic
black appearance of this material.
\\
The peak positions appear to be doping independent, but
the striking observation is that
the {\em intensity} of the peaks depends strongly
on doping. In particular, the measured intensity of the 1 eV peak
for the {\em a} axis is directly proportional to $1-x$
(Fig. \ref{intens}). The 1 eV peak
for the {\em b} axis shows also a decrease upon doping.

\section{Main elements of the electronic structure}
\begin{figure}
\centerline{\epsfig{figure=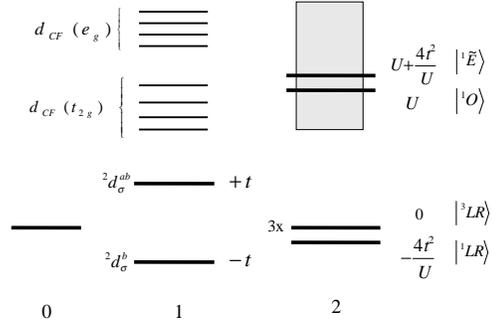,width=7cm,clip=}} \caption{
Diagram of the many-body eigenstates of a VOV rung, occupied with 0,
1 and 2 electrons, obtained within the Heitler-London model for $U=4 eV$
$\Delta=0.8 eV$, and and $t_{\perp}=0.3 eV$. }
\label{rungstates}
\end{figure}

\begin{figure}
\centerline{\epsfig{figure=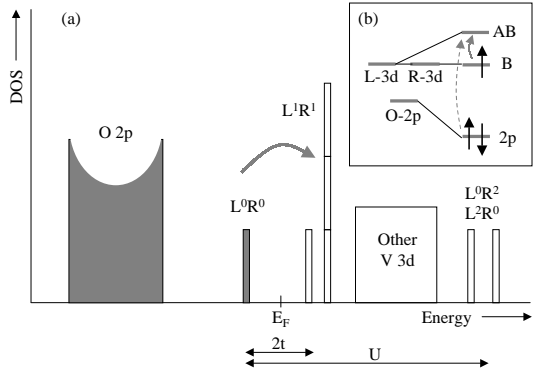,width=7cm,clip=}} \caption{
Occupied and unoccupied density of states obtained within the
Heitler-London model for $U=4 eV$ (a), and the schematic diagram
of the relevant energy levels of the V$_2$O clusters forming a
rung (b). L and R refer to orbitals associated to the left and
right V sites, respectively. }
\label{navodos}
\end{figure}

Before entering the interpretation of the data, we need
to discuss in some more detail the main elements of the
electronic structure of these compounds.
The basic building block of the crystal structure of
$\alpha^{\prime}$-NaV$_2$O$_5$ is formed by VOV dimers. These
dimers form the rungs of quasi one-dimensional ladders. The
V-ions forming the rungs are bonded along the ladder direction
via oxygen ions.
\\
The backbone of the electronic structure is formed by
the oxygen $2p$ and V$3d$ states. Photoelectron
spectroscopy\cite{kobayashi} has provided crucial
information on the occupied electronic levels: The
oxygen $2p$ states have the lowest energy (the highest
binding energy). They form a band about 4 eV wide,
which is fully occupied. The occupied part of the
V $3d$ states is located about 3 eV above the top
of the oxygen bands. Due to ligand field splittings the
V $3d$ manifold is spread over a range of at least 3 eV.
The $3d_{xy}$ is of the V$^{4+}$ is occupied with one
electron. The unoccupied $d_{xz}$ and $d_{yz}$ levels have
an energy at least 1 eV higher. These in turn are located
about 2 eV below the $d_{x^2-y^2}$ and the $d_{z^2}$
levels\cite{smolinski}. 
The relevance of the O 2p bands is that they
provide a path for virtual hopping processes between the V-sites.
The coupling between V-sites is
through virtual hopping via $\pi$-bonded O $2p_y$ states,
indicated schematically in Fig.\ref{navodos}b. The
effective hopping parameter
between V-sites is $t_{\perp}\approx -0.3 eV$ on the same rung,
and $t_{\perp}\approx -0.2 eV$ along the 
legs of the ladder\cite{phorsch,nishi,cuoco}.
The number of electrons is one per pair of V-atoms. Approaching
the ladders as a linear array of rungs, weakly coupled along the
direction of the ladder, results in a model of electrons occupying
a narrow band of states formed by the anti-symmetric combination
of the two V 3d-states forming the rungs, hereafter referred to as
V-V bonding levels.
\\
Hence, we see that the basic building block are the
pairs of V$3d$ states, together forming the rungs of the ladders.
The essential charge and spin degrees of freedom of a single rung
are identical to the Heitler-London model of the H$_{2}^{+}$ ion,
with the V$3d_{xy}$ states playing the role of the H $1s$
states\cite{phorsch}. The relevant Hamiltonion is
\begin{equation}
\begin{array}{l}
 H = t_{\perp}\sum_{\sigma}\left\{
   d_{L\sigma}^{\dagger}d_{R\sigma}+
   d_{R\sigma}^{\dagger}d_{L\sigma}\right\} \\
   + \frac{\Delta}{2}\sum_{\sigma}\left\{n_{L\sigma}-n_{R\sigma}\right\}
   + U\left\{n_{L\uparrow}n_{L\downarrow}+n_{R\uparrow}n_{R\downarrow}\right\}
\end{array}
\label{hamil}
\end{equation}
where $d_{L(R),\sigma}^{\dagger} $ creates an electron in the lefthand
(righthand) $d_{xy}$ orbital on the rung, and $\sigma$ is the spin-index.
The bias potential $\Delta$ between the two V-sites accounts for a
possible left/right charge imbalance\cite{damascelli}. The most relevant
states for the ground state are $d_{L,xy}$ and $d_{R,xy}$. Pure
NaV$_2$O$_5$ contains one electron per rung in the ground state.
Important in the present discussion are the eigenstates and
energies of a rung with 0, 1 or 2 electrons. The eigenstates and
energies are listed in Table \ref{states}.
\\
In Fig. \ref{rungstates} the level diagram is displayed. In this representation
$N_e=0(2)$ corresponds to the one electron removal(addition) states,
for noninteracting electron picture indicated as the "occupied" ("empty") states.
In Fig.\ref{navodos} we display the same information represented as the
the one electron removal and addition spectral function. For
non-interacting electrons this represents the occupied (left) and
unoccupied (right) states.
\\
In the absence of electron-electron interactions,
here represented by the on-site Hubbard repulsion parameter $U$,
the Fermi energy would be located in the middle of the bonding
band, resulting in a metallic conductor.
In Figs.\ref{rungstates} and \ref{navodos} we have adopted
$U=4$ eV. The model now predicts a gap of order
$E_{CT}\simeq$1 eV. The Fermi energy is located within
this gap. The fact that these materials are insulating
therefore is associated with the large on-site
Hubbard interaction.
\\
The excitation of an electron across the gap
involves a change of occupancy of {\em two}
of the rungs: The final state has one empty  and one
doubly occupied rung. It is important in this context, that
the two electrons $|^{3,1}LR>$ are in a {\em correlated}
state: In the limit  $U\rightarrow\infty$ one electron is located
on the lefthand V-atom and the other on the righthand V-atom.
\section{Discussion of the experimental spectra}
The 0.9 eV peak marks the fundamental gap of the optical
spectrum.  The interpretation of this peak is still subject
of a scientific controversy. Several interpretations have been
put forward
\begin{enumerate}
\item
Transitions between linear combinations of
V 3d$_{xy}$-states of the two V-ions forming the
rungs\cite{smolinski,phorsch,damascelli}. In Refs.\cite{smolinski}
and \cite{phorsch} even and odd combinations were considered.
The 0.9 eV peak in $\sigma_{a}(\omega)$ (peak A) would then correspond
to the transition from V-V bonding to antibonding combinations on the same
rung\cite{damascelli} (Fig.\ref{navodos}b). In Ref.\cite{damascelli}
this model was extended to allow lop-sided linear combinations of the
same orbitals, so that the 0.9 eV peak is a transition between
left- and right-oriented linear combinations.
\item
On-site d-d transitions between the crystal field
split levels of the V-ions\cite{golubchik}. Because
V$^{5+}$ has no occupied 3d-levels, such processes involve
the V$^{4+}$ ions.
\item
Transitions between the on-rung V 3d$_{xy}^+$ bonding combination
and final states of $d_{xz}$ and $d_{yz}$ character\cite{long}.
\end{enumerate}
Optical transitions having values below 2 eV
were also seen in V$_6$O$_{13}$ and VO$_2$. In  V$_2$O$_5$ they have
very small intensities, and were attributed to defects \cite{kenny}.
The last two assignments are motivated by the fact that in
$\alpha^{\prime}$-Na$_{1-x}$Ca$_x$V$_2$O$_5$
the optical selection rules allow on-site d-d transitions
by virtue of the low point-symmetry at the V-sites.
To determine which one of these assignments is true, we have
measured the doping dependence of the 0.9 eV peak in
$\sigma_{a}(\omega)$ in Ca-substituted
$\alpha^{\prime}$-NaV$_2$O$_5$ (Fig.\ref{navoca}a). Because Ca is
divalent, substituting Na with Ca has the effect of increasing the
average density of V$^{4+}$ ions.
\\
In the local d-d scenario the intensity
of the on-site V 3d-3d transitions would be proportional to the
density of V$^{4+}$ ions. As a result the intensity of on-site d-d
transitions increases upon substituting Na with Ca.
In Fig. \ref{intens} we display the experimentally observed
doping dependence of the intensity together with the theoretical
expectation within this scenario. Clearly
the experimental intensity of the 0.9 eV peak in Fig. \ref{navoca}
behaves opposite to the expected behaviour of $dd$ transitions.
This definitely rules out item number 2 of the above list.
\\
This also rules out item number 3 presented in Ref.\cite{long}:
if the transition from  the V 3d$_{xy}^+$ bonding combination
to the $d_{xz}$ and $d_{yz}$  orbitals involves mainly transitions among orbitals
at the same site, the same argument as for item 2 applies. If
it involves mainly transitions between molecular orbitals formed
by different sites on the same rung, the transition from the 
V 3d$_{xy}^+$ bonding combination to the antibonding V 3d$_{xy}^-$
would still be the dominant transition.
\\
To explain the intensity decrease of about x\% upon substituting
x\% of the Na ions with Ca (see Fig.\ref{intens}), let us have a
look at the many-body nature of the ground state and the excited
state properties of NaV$_2$O$_5$. The rungs which are occupied
with two electrons due to the Ca-doping will be in the many-body
ground state (see Table \ref{states}) $|^1\tilde{LR}\rangle$. Due
to the dipole selection rules, the optical excitation with the
electric field along the rung-direction is
$|^1\tilde{LR}\rangle\rightarrow|^1\tilde{O}\rangle$. The energy
to make this excitation is
$\frac{1}{2}U+\sqrt{U^2/4+4t_{\perp}^2}$. Hence the effects of Ca
doping are (i) to remove the peak at $E_{CT}\approx 1 eV$ for the rungs
receiving the extra electron, and (ii) to place a new peak at an energy
$U\approx 4 eV$. Hence, the observed x\% decrease of intensity of the
$|B\rangle\rightarrow|A\rangle$ transition peak for the x\% Ca
doped sample is in excellent quantitative agreement with the
expected value.
\\
Using the tightbinding f-sum rule
$
\int\sigma(\omega)d\omega=
 (e d/\hbar)^2 \pi t_{\perp}/(2V)
  \langle d_{L\sigma}^{\dagger}d_{R\sigma}+\mbox{HC}\rangle
\label{tbfsum}
$
the {\em intensity} of the
$|^1\tilde{LR}\rangle\rightarrow|^1\tilde{O}\rangle$ peak relative
to the 0.9 eV peak of the singly occupied rungs is (assuming $\Delta=0$
for a doubly occupied rung):
\begin{equation}
 \frac{I(0)+I(2)}{2I(1)}=\frac{\alpha_K\beta_K}{uv}=\sqrt{\frac{1}{1+(U/4t)^2}}
 \approx \frac{4t_{\perp}}{U}
\end{equation}

With the parameters relevant to NaV$_2$O$_5$ this implies that the
$|^1\tilde{LR}\rangle\rightarrow |^1\tilde{O}\rangle,|^1\tilde{E}\rangle$
transitions have factor 2-4 smaller spectal weight than
the $|B\rangle\rightarrow|A\rangle$ transition.
Hence we conclude that only the assigment of item number 1 
is consistent with our data:
The 1eV peak in $\sigma_a(\omega)$ is
the on-rung $|^2B_{\sigma}>\rightarrow|^2A_{\sigma}>$ transition
with an excitation energy $E_{CT}\equiv\sqrt{4t_{\perp}^2+\Delta^2}$.
\\
The 1.1 eV peak in $\sigma_{b}(\omega)$ (peak B)
involves transitions between neighboring rungs
along the ladder. In the non-interacting model ($U=0$)
this would correspond to a Drude-Lorentz optical conductivity
centered at $\omega=0$, with a spectral weight
$\int_0^{\infty}\sigma_{b}(\omega)d\omega= (e/\hbar)^2
t_{\parallel} \pi b (2ac)^{-1}$.
As a result of the correlation gap in the density of states,
indicated in Fig. \ref{navodos}a, the optically
induced transfer of electrons
between neighboring rungs results in a final state with one
rung empty, and a neighboring rung doubly occupied, in
other words, an electron hole pair consisting of a hole in
the band below E$_F$, and an electron in the empty state above
E$_F$ indicated in Fig. \ref{navodos}a.
This corresponds to the process
\begin{equation}
 2|^2B_{\sigma}> \rightarrow |^{3,1}LR> + |0>
 \label{process}
\end{equation}
Note that the final state
wavefunction is qualitatively different from the
on-rung bonding-antibonding excitations considered above,
even though the excitations are close in energy: it involves one
rung with no electron, and a neighboring rung with one electron
occupying each V-atom.
The energy of this process is approximately $2t_{\perp}+\delta V$, where
$\delta V$ represents the increase in Coulomb interaction by bringing two
electrons together on the same rung. Since the distance betweeen the electrons
changes from about 5.0$\AA$ to 3.4$\AA$, and taking into account a
screening factor $\epsilon\approx 6$, we estimate that $\delta V\simeq 0.2 eV$.
\\
This value of $\delta V$ corresponds closely to the
difference in peakpositions
along the $a$ and $b$ directions. According to this interpretation the
absorption at 1.1 eV along $b$ corresponds to the creation of a free
electron and hole, capable of carrying electrical currents.
The on-rung excitation at 0.9 eV along the $a$-direction
is a localized (charge neutral) excitation, in other words an exciton. In
this case the energy of the exciton
involves the states of a single electron only, whereas the free
carrier states involve many-body interactions.
\\
Doping with Ca creates doubly occupied rungs, whose ground
energy is not  $2|^2B_{\sigma}>$  but  $|^1\tilde{LR}\rangle$.
Consequently, the electrons on these rungs will not be involved in 
the processes of Eq. \ref{process}, 
thus decresing the intensity of the B peak upon doping,
as seen from Fig. \ref{navoca}b. 
\\
The room temperature crystal structure has four V-atoms per unit
cell, organized in ladders with up and down oriented apical
oxygens alternating along the a-direction, resulting in a double
degeneracy of the electronic states discussed above. The coupling
between adjacent ladders lifts the degeneracy of these states,
resulting  in a "Davidov" splitting of the peaks A and B. This can
create the two additional "shoulders" in $\sigma(\omega)$ at 1.4
and 1.7 eV for peak A and B respectively.
\\
With ARPES\cite{kobayashi} it has been observed that
an energy of 3 eV separates the V$3d$ band from the O$2p$.
We therefore attribute the peaks at 3.3 eV in
$\sigma_{a}(\omega)$ and the peak at 4 eV in $\sigma_{b}(\omega)$
to transitions of the type
\begin{equation}
 |^2B_{\sigma}> \rightarrow |\underline{2p}^{1}LR>
\end{equation}
where the $\underline{2p}$ hole is located on the oxygen on the same
rung for peak A, and inbetween the rungs for peak B.
This is further supported by previous optical measurements on
V$_2$O$_5$ \cite {kenny} showed a peak at about 3 eV. In
V$_2$O$_5$ all V-ions have a formal V$3d^0$ configuration,
hence the 3 eV peak can not be attributed to d-d transitions. However,
the O$2p\rightarrow$V$3d$ transitions should appear at approximately
the same photon energy as in NaV$_2$O$_5$, which further supports
our assigment of the 3 eV peak in NaV$_2$O$_5$ to
O$2p\rightarrow$V$3d$ transitions.
\section{Conclusions}
In conclusion, we have measured the dielectric function along the
{\em  a} and {\em  b} axes of Ca$_x$Na$_{1-x}$V$_2$O$_5$ for x=0,
0.06, 0.15 and x=0.20. The 0.9 eV peak in $\sigma_{a}(\omega)$ was
shown experimentally to be a bonding-antibonding transition inside
the V$_2$O rung and not a vanadium d-d transition due to crystal
field splitting. We identified the 3.3eV peak in
$\sigma_{a}(\omega)$ as the transition from the oxygen orbitals to
the antibonding one. This strongly supports the notion, previously
expressed in Refs \cite{smolinski,phorsch,damascelli} that
NaV$_2$O$_5$ is an insulator due to a combination of three
factors: A crystal field splitting, an on-site Hubbard
interaction, and an on-rung bonding-antibonding splitting of the
two V$3d_{xy}$ orbitals, each of which is large compared to the
inter-rung hopping parameter.

\section{Acknowledgements}
We like to thank H. Bron for his assistence with the chemical analysis of the 
crystals, and prof. J.T.M. de Hosson for making available the microprobe 
equipment. This investigation was supported by the Netherlands Foundation for
Fundamental Research on Matter (FOM) with financial aid from the
Nederlandse Organisatie voor Wetenschappelijk Onderzoek (NWO).
\newpage
\onecolumn
\widetext

\begin{table}
\begin{tabular}{lll}
N$_e$& State vector & Energy \\
\hline
$0$& $|0\rangle$&$0$  \\
\hline
$1$& $|^2B_{\sigma}\rangle=
    (ud_{L\sigma}^{\dagger}+vd_{R\sigma}^{\dagger})|0\rangle$ &
                                            $-\frac{1}{2}E_{CT}-E_F$ \\
   & $|^2A_{\sigma}\rangle=
    (vd_{L\sigma}^{\dagger}-ud_{R\sigma}^{\dagger})|0\rangle$ &
                                           $+\frac{1}{2}E_{CT}-E_F$ \\
\hline
$2$& $|^1\tilde{LR}\rangle=\alpha_K|^1LR\rangle+\beta_K|^1E\rangle$
     & $-K-2E_F$ \\
   & $|^3LR_1\rangle=d_{L\uparrow}^{\dagger}d_{R\uparrow}^{\dagger}|0\rangle$
     & $-2E_F$ \\
   & $|^3LR_0\rangle=\sqrt{\frac{1}{2}}
        (d_{L\uparrow}^{\dagger}d_{R\downarrow}^{\dagger} +
         d_{L\downarrow}^{\dagger}d_{R\uparrow}^{\dagger})|0\rangle$
     & $-2E_F$ \\
   & $|^3LR_{-1}\rangle=d_{L\downarrow}^{\dagger}d_{R\downarrow}^{\dagger}|0\rangle$
     & $-2E_F$ \\
   & $|^1\tilde{O}\rangle=\alpha_{\Delta}|^1O\rangle
                   +\beta_{\Delta}\alpha_K|^1E\rangle
                   -\beta_{\Delta}\beta_K|^1LR\rangle$
     & $U+K/2-\sqrt{(K/2)^2+\Delta^2}-2E_F$ \\
   & $|^1\tilde{E}\rangle=\alpha_{\Delta}\alpha_K|^1E\rangle
                   -\alpha_{\Delta}\beta_K|^1LR\rangle
                   -\beta_{\Delta}|^1O\rangle$
     & $U+K/2+\sqrt{(K/2)^2+\Delta^2}-2E_F$ \\
\hline
Definitions&&\\
\hline
 &$\frac{u}{v}\equiv
    \sqrt{1+\left[\Delta/2t_{\perp}\right]^2}
    +\left[\Delta/2t_{\perp}\right]$
     &$E_{CT}\equiv\sqrt{4t_{\perp}^2+\Delta^2}$\\
 &$|^1LR\rangle\equiv\sqrt{\frac{1}{2}}
         (d_{L\uparrow}^{\dagger}d_{R\downarrow}^{\dagger} -
         d_{L\downarrow}^{\dagger}d_{R\uparrow}^{\dagger})|0\rangle$
 &$K\equiv\sqrt{U^2/4+4t_{\perp}^2}-U/2$\\
 &$|^1E\rangle\equiv\sqrt{\frac{1}{2}}
       (d_{L\uparrow}^{\dagger}d_{L\downarrow}^{\dagger} +
        d_{R\uparrow}^{\dagger}d_{R\downarrow}^{\dagger})|0\rangle$
 &$\frac{\alpha_K}{\beta_K}\equiv
    \sqrt{1+\left[U/4t_{\perp}\right]^2}+\left[U/4t_{\perp}\right]$\\
 &$|^1O\rangle=\sqrt{\frac{1}{2}}
     (d_{L\uparrow}^{\dagger}d_{L\downarrow}^{\dagger}
     -d_{R\uparrow}^{\dagger}d_{R\downarrow}^{\dagger})|0\rangle$
 &$\frac{\alpha_{\Delta}}{\beta_{\Delta}}\equiv
        \sqrt{1+\left[K/2\Delta\right]^2}+\left[K/2\Delta\right]$
\end{tabular}
\caption{Eigenstates and energies of a rung with N$_e$=0, 1 and 2 electrons.
The N$_e$=2 state vectors and energies were derived for the limit $\Delta \ll U$.}
\label{states}
\end{table}

\begin{references}
\bibitem{note1}
 Because $\sigma(\omega)$ is a direct measure of the spectral
 weight in relation to the $f$-sum rule, we use $\sigma(\omega)$ as
 a representation of $\mbox{Im}\epsilon(\omega)$ throughout this
 paper.
\end{references}
\end{document}